\def\thetacl{\theta_{\rm cl}}
\def\thetaclo{\theta_{\rm cl, 0}}
\def\Abar{\overline{A}}
\def\Abarfourty{\overline{A}_{40}}
\def\xmax{x_{\rm max}}
\def\ffill{f_{-2}}
\def\rcl{r_{\rm cl}}
\def\Mcl{M_{\rm cl}}
\def\mcl{m_{\rm cl}}
\def\ncl{n_{\rm cl}}
\def\Nlos{N_{\rm cl}^{\rm l.o.s.}}
\def\Plos{P_{\rm cl}^{\rm l.o.s.}}

\def\taufecl{\tau_{\rm Fe, cl}}
\def\Xfe{X_1^{\rm Fe}}

\def\tion{t_{\rm ion}}
\def\trec{t_{\rm rec}}
\def\Zeff{Z_{\rm eff}}

\def\ls{\lower4pt\hbox{${\buildrel < \over \sim}$}}
\def\gs{\lower4pt\hbox{${\buildrel > \over \sim}$}}

\documentclass[12pt, preprint]{aastex}

\slugcomment{Accepted for publication in {\it The Astrophysical Journal}}

\shorttitle{Transient absorption features in GRBs}
\shortauthors{B\"ottcher et al.}

\begin{document}

\title{Transient Absorption Features in GRBs and Their Implications
for GRB Progenitors}

\author{M. B\"ottcher\footnote{Chandra Fellow}}
\affil{Department of Physics and Astronomy, Rice University, MS 108, \\
6100 S. Man Street, Houston, TX 77005 - 1892, USA}
\email{mboett@spacsun.rice.edu}

\and

\author{C. L. Fryer\footnote{Feynman Fellow}}
\affil{Theoretical Astrophysics, Los Alamos National National Laboratory,\\
Los Alamos, NM 87545, USA}
\email{fryer@lanl.gov}

\and

\author{C. D. Dermer}
\affil{E. O. Hulburt Center for Space Research, Code 7653,\\
Naval Research Laboratory, Washington, DC 20375-5352}
\email{dermer@gamma.nrl.navy.mil}

\begin{abstract}
The recent detection of a transient absorption feature in the
prompt emission of GRB~990705 has sparked multiple attempts to
fit this feature in terms of photoelectric absorption or
resonance scattering out of the line of sight to the observer.
However, the physical conditions required to reproduce the
observed absorption feature turn out to be rather extreme
compared to the predictions of current GRB progenitor models.
In particular, strong clumping of ejecta from the GRB progenitor
seems to be required. Using detailed 3D hydrodynamic simulations
of supernova explosions as a guideline, we have investigated 
the dynamics and structure of pre-GRB ejecta predicted in 
various GRB progenitor models. Based on our results, combined 
with population synthesis studies relevant to the He-merger
model, we estimate the probability of observing X-ray 
absorption features as seen in GRB~990705 to $\ll 1$~\%. 
Alternatively, if the supranova model is capable of producing
highly collimated long-duration GRBs, it may be a more promising
candidate to produce observable, transient X-ray absorption
features. 
\end{abstract}

\keywords{galaxies: active --- gamma-rays: theory}  

\section{Introduction}

Recently, \cite{amati00} have reported the detection of a 
transient absorption feature, consistent with a redshifted 
Fe~K edge at the redshift of the burst, $z = 0.86$ (see
\cite{lazzati01}), in the prompt X-ray emission from GRB~990705. 
This is the second time (after GRB~980329; \cite{frontera00}) 
that X-ray spectroscopy during the prompt phase of a GRB revealed 
evidence for excess soft X-ray absorption above the Galactic 
hydrogen column, and the first time that evidence for a time
dependence of such absorption has been found. (Here, we do not 
consider the possible absorption features detected by Ginga in 
a few cases [see, e.g., \cite{yoshida92}] at higher energies 
which were inconsistent with being due to photoelectric absorption 
or other atomic processes.) The transient nature of the absorption 
feature was rather significant: The absorption edge was detected 
at the $\sim 12 \sigma$ level in observing interval B (with its
depth being consistent with the less significant detection in
interval A), while there was $< 2 \sigma$ evidence for such an
absorption edge in observing interval D (and all following intervals), 
when the upper limit on its depth was inconsistent with the value 
measured during interval B at the $\ge 5 \sigma$ level. This has 
been interpreted as evidence for photoionization of the absorber 
by the prompt burst emission \citep{amati00,boettcher01a,boettcher01b}. 
In contrast to the non-transient excess absorption in GRB~980329
\citep{frontera00}, which does not yield very specific information
about the location of the absorbing material, the transient nature
of an X-ray absorption feature allows rather detailed diagnostics
of the relevant physical time scales and the geometry of the
GRB environment.

If the observed absorption feature is due to photoelectric 
absorption at the Fe K edge, then it is natural to try to 
interpret the duration of the occurrence of this feature 
as the photoionization time scale, assuming that the absorbing 
material is not dense enough for recombination to be efficient.
The expected X-ray absorption signatures of such a scenario had 
been investigated previously by \cite{boettcher99} who had 
concluded that such features are either transient on the prompt
GRB time scale, or remain virtually unchanged throughout the
prompt and afterglow phase. However, the application of this idea
to the specific case of GRB~990705 by \cite{boettcher01a} revealed
that either an implausibly large amount of iron, concentrated very 
close to the GRB, would be required, or the iron would have to
be distributed very anisotropically around the burst source.

As an alternative, \cite{boettcher01a,boettcher01b} have investigated 
an environment containing small, dense clumps of iron-enriched 
material. In such an environment, photoionization can be balanced by 
rapid recombination, and the duration of the absorption feature 
can be identified with the Compton heating time scale of the 
absorber by the prompt GRB emission, or with the sweep-up time scale 
of the absorbing material by the relativistic blastwave responsible 
for the GRB and afterglow emission. This scenario can produce the
observed absorption feature in GRB~990705 with a reasonable amount
of iron, quasi-isotropically distributed around the GRB source, but
required that the clumps containing this material be very dense and 
exhibit a rather extreme degree of clumping. 

Finally, \cite{lazzati01} have suggested that the absorption 
feature could be a blue-shifted resonance scattering feature in an 
inhomogeneous high-velocity outflow. This requires that the absorbing 
material has a systematic average outflow velocity of $v_0 \sim 0.13$~c
and a velocity dispersion $\Delta v \sim v_0$. While this scenario can
slightly reduce the required amount of iron and degree of clumping,
it requires a kinetic energy of $\sim 10^{53} \, (\Omega / 4 \pi)$~ergs 
in the directed outflow, which has to be ejected several months prior 
to the GRB.

The parameter estimates resulting from all three of these basic 
scenarios are summarized in Table \ref{parametertable}. Those
estimates are rather generic and independent of specific progenitor
scenarios. In this paper, we are tying parameter estimates pertaining 
to the transient absorption feature in GRB~990705 to specific GRB
progenitor models. This is done by deducing constraints on the 
supernova explosion which is likely to have happened prior to 
the GRB, ejecting the material responsible for the transient 
absorption feature. In \S \ref{models} we briefly review the 
viable GRB progenitor models which are generally capable of
producing a significant concentration of high-Z material close
to the GRB prior to the actual GRB event. The degree of clumping
of matter ejected in supernova explosions will be discussed in
\S \ref{clumping}. Model-specific estimates of the parameters
pertaining to the pre-ejected material will be calculated in
\S \ref{absorptionfeatures}. We will discuss the implications
of our results in \S \ref{discussion}.

\section{\label{models}GRB models with pre-GRB supernovae}

With most long-duration GRB progenitors it is hard to produce a
sizable amount of iron within 1~pc of the GRB engine.  For most
collapsar progenitors the only source of iron prior to collapse would
be in the massive star's wind.  Unless these stars are extremely metal
rich, the total mass in iron in these winds will be much less than
0.1~$M_{\odot}$.  Unfortunately, metal rich massive stars lose too much
mass to winds and will probably not form collapsars \citep{fryer01}.

The evolutionary scenario of He-merger GRBs provides a more likely,
although still rare, source of iron within $\sim 1$~pc of the GRB engine.
Recall that the formation process of a He-merger GRB requires a binary
system whose primary evolves to collapse and produces a compact
remnant.  Normal supernova explosions eject from 0.002-0.3~$M_{\odot}$ 
of $^{56}$Ni \citep{turatto98,schmidt94} into the region
surrounding the binary system and this $^{56}$Ni ultimately decays
into iron.  It is possible that the primary produces an abnormal
supernova like supernova 1998bw which then may eject as much as a
solar mass of $^{56}$Ni or more \citep{sollerman00,germany00}.
When the primary remnant eventually merges with its companion, 
a GRB is formed, exploding into this iron rich region.

The difficulty lies in limiting the iron to a region within $\sim 1$~pc 
of the binary system until the merger occurs and a GRB is produced 
(see the values of $x$ in Tab. \ref{parametertable}).
Because the $^{56}$Ni is produced in the inner layers of an exploding
supernova and a large fraction of $^{56}$Ni is ejected at low
velocities ($\sim$ 1000~km~s$^{-1}$).  Although these velocities are 
relatively low, it takes only 1000~yr for this ejecta to travel 1~pc.
Even if a sizable amount of $^{56}$Ni is ejected with lower
velocities, winds from the companion star (which tend to be massive
for most He-merger GRBs) will blow the $^{56}$Ni away from the 
binary at roughly 1000~km~s$^{-1}$ and it is difficult to increase
the maximum allowed time delay between primary supernova and GRB 
outburst (to be consistent with the previous parameter estimates
for the absorber in GRB~990705) by less than 1000~yr.

With the standard formation scenario of He-merger GRBs, such small
delays between supernova explosion and GRB outburst require that the
secondary star evolve off the main sequence a scant 1000~yr after the
primary supernova explosion.  This would require extreme fine tuning
of the masses of the two stars and the likelihood of such an occurrence
would be less than 0.01~\%.

However, an alternate formation scenario may be more plausible. If
the secondary evolves off the main sequence before the primary
explosion, a common envelope phase will occur where two helium cores
tighten their orbits as they inspiral within a hydrogen common
envelope.  Typically, after the hydrogen envelope is ejected, it
leaves behind two tightly orbiting helium cores.  When the primary's 
helium core collapses, the system may be disrupted by neutron star 
kicks, it may remain bound, or the neutron star may be placed in 
such a close orbit that it quickly merges (within 1000~yr) with its 
helium star companion.  The latter case, where the neutron star 
merges with its helium companion, produces a He-merger GRB 
surrounded by the iron ejecta of the primary's supernova. Using the 
binary population synthesis code developed in \cite{fryer98},
we have calculated the rate of these quick He-merger GRBs for
a range of population synthesis parameters.  The fraction of
quick mergers lies somewhere between 0.5-2.5~\%.  For most of the
double helium cores which actually form He-merger GRBs, there is
typically a delay of 10-100 years from the supernova explosion to 
the GRB outburst.

SN - GRB delays significantly below $\sim 10$~yrs could occur 
in the supranova model of \cite{vs98}, where the supernova
explosion of a fast rotating, massive star leaves behind a
rapidly spinning, supra-massive neutron star which implodes
several months to several years later to form a GRB. The time
delay between the SN and the GRB will then be determined by 
the spin-down time, which can be estimated as $t_{sd} \sim 
10 \, (j/0.6) \, \omega_4^{-4} \, B_{12}^2$~yr, where $j$ 
is the dimensionless angular momentum of the neutron star, 
$\omega_4$ is its angular velocity in units of 
$10^4$~s$^{-1}$, and its surface magnetic field is 
$B = 10^{12} \, B_{12}$~G. Detailed simulations of collapsing 
neutron stars \citep{ruffert96,fw98} seem to indicate that, 
in contrast to Vietri \& Stella's original suggestion, this 
process ejects a rather large amount of baryonic material and 
has too little energy to produce a GRB. However, strong beaming 
of a relativistic outflow along the rotational axis may overcome 
the latter problem so that we will consider the supranova model 
as a conceivable alternative in this paper, although more detailed 
simulations are needed in order to address the concerns mentioned 
above. 

\section{\label{clumping}Clumping of ejecta}

Supernova 1987A surprised astronomers by giving strong evidence 
that the $^{56}$Ni ejected in the explosion was mixed into the 
outer layers of the star \citep{dotani87,itoh87,sunyaev87,matz88}.
This $^{56}$Ni fragmented into high velocity ``bullets'' 
\citep{spyromilio88} which could potentially be the clumps 
necessary to explain the absorption feature in GRB~990705.  
It is believed that Rayleigh-Taylor instabilities which arise 
when the shock moves through the composition layers of the star 
produce this mixing and lead to clumps of $^{56}$Ni on size 
scales from 1 to 5 degrees (see \cite{hw94}, \cite{kifonidis00}
and references therein). Spatially resolved spectroscopy of
Galactic SNRs by the {\it Chandra} X-ray Observatory (e.g., 
\cite{hughes00,hwang00}) confirmed the tendency of element 
mixing and clumping of the high-z material into dense blobs 
in SNRs.

\begin{figure}
\plotone{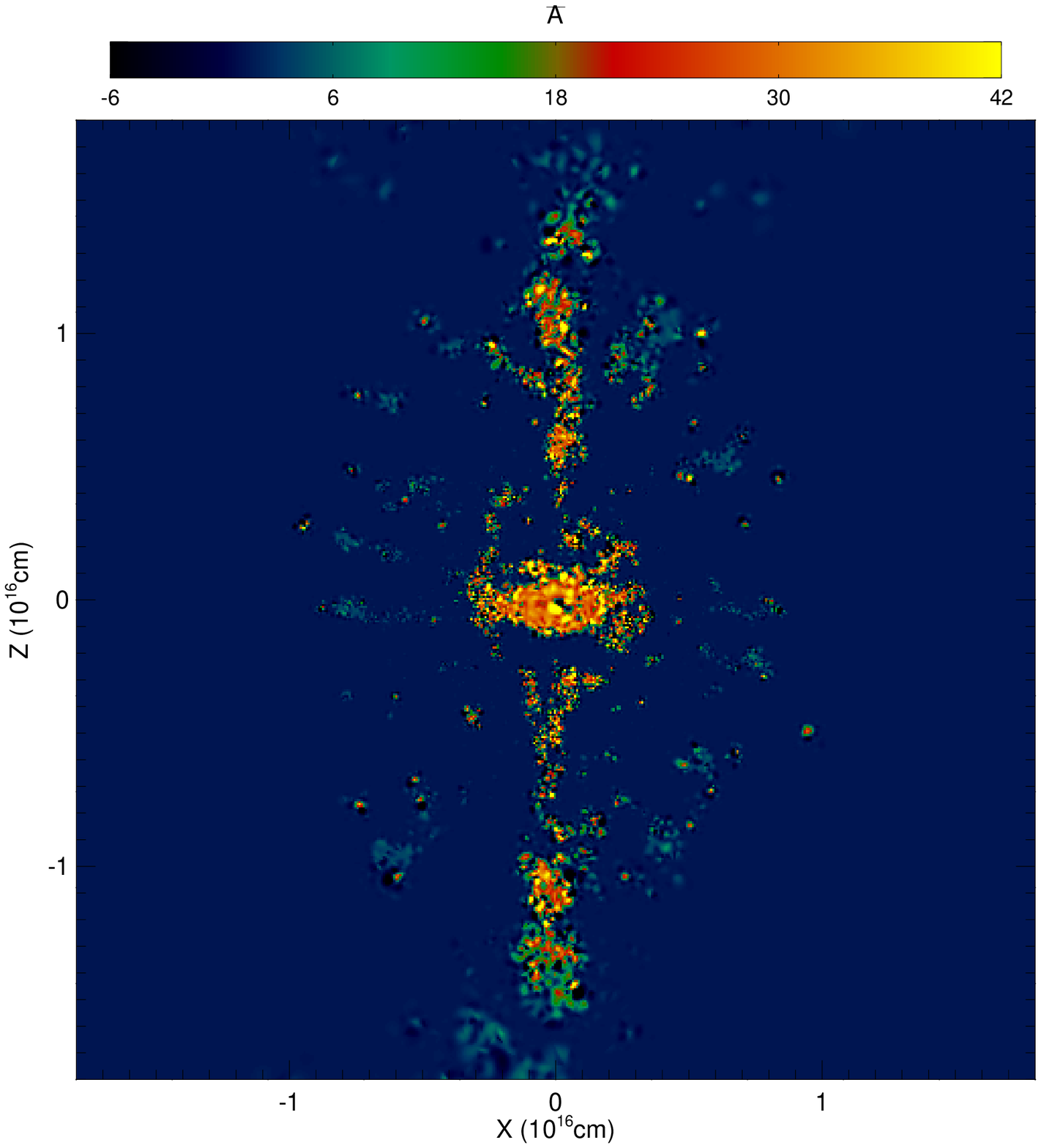}
\caption{3D simulation of a supernova explosion of a $15 \, M_{\odot}$ 
star (for details see \cite{fhw01}). The figure shows the distribution 
of the ejecta, color-coded by the average atomic weight, 1~yr after the 
SN explosion.}
\label{snclumps}
\end{figure}

This is in agreement with detailed 3D hydrodynamic simulations 
of SN explosions as described in detail in \cite{fhw01}. 
Fig. \ref{snclumps} illustrates an example of the slightly 
asymmetric supernova explosion of a $15 \, M_{\odot}$ star.
The figure shows the distribution of the ejecta, color-coded
by the average atomic weight, 1~yr after the SN explosion.
It shows that most of the heavy elements are concentrated in
dense clumps of angular size scale $\sim 1$ -- a few degrees,
and are concentrated around $\overline x \sim 5 \times 10^{15}$~cm
at that time.

In the following, we will use the results the parameters
of the supernova ejecta derived from this example to estimate
the expected spatial distribution and, consequently, the expected
time-dependent absorption features produced in pre-GRB
supernova ejecta in the He-merger and the supranova scenarios.
For this purpose, we assume that the metal-rich, dense clumps
are not decelerating significantly, so that their distance
from the center of the explosion scales linearly with the
SN-GRB time delay $\Delta t$. Different GRB progenitor models 
will then differ primarily by $\Delta t$ and by the amount of 
mass in the clumpy ejecta, $M_{\rm cl}$, which might be correlated 
with the mass of the progenitor. We are thus interested in the 
location of model systems in the $M_{\rm cl}$-$\Delta t$ plane 
which are capable of reproducing the absorption feature 
seen in GRB~990705.

As mentioned above, the directed velocity of clumps containing
an overabundance of heavy elements, might be $v_0 \ls 
10^9$~cm~s$^{-1}$. According to the analysis of \cite{lazzati01},
this would produce too narrow an absorption feature at a too low
centroid energy, if the dominant absorption mechanism were 
resonance scattering out of the line of sight by hydrogen-like 
iron. For this reason, we will concentrate in the following on 
photoelectric absorption as the dominant absorption mechanism.

\section{\label{absorptionfeatures}Absorption features from
pre-GRB supernova ejecta}

As motivated above, we assume that 
most of the iron is concentrated in 
metal-rich clumps on angular scales of 
$\thetacl \sim \thetaclo$~degrees, 
with $\thetaclo \sim 1$ -- 5, within which
the average atomic weight $\Abar \sim 40 \, 
\Abarfourty$ may exceed $\Abarfourty \sim 1$. 
Those clumps are filling a fraction $f = 1 \,
\ffill$~\% of the volume within a typical
size scale of $\xmax = 10^{16}$~cm 1~yr after 
the supernova explosion. As a function of the
time delay $\Delta t = 1 \, t_y$~yr, between
the SN explosion and the GRB, the spatial distribution 
of clumps will extend out to $\xmax \sim 10^{16} \, 
t_y$~cm. Assuming that most of the clumps will be 
concentrated around $\overline x \sim \xmax/2$, this 
yields a typical size scale of such clumps as $\rcl 
\sim 9 \times 10^{13} \, t_y \, \thetaclo \; {\rm cm}$. 
For a total supernova ejecta mass concentrated
in clumps, $\Mcl = \mcl \, M_{\odot}$, we then
find the average density of nuclei in the clumped
ejecta to be
\begin{equation}
\ncl \sim 2.9 \times 10^{10} \, {\mcl \over \Abarfourty 
\, \ffill \, t_y^3} \; {\rm cm}^{-3}. 
\label{ncl}
\end{equation}
We would then expect to have $\Nlos \sim 0.6 \, \ffill / 
\thetaclo$ clouds located by chance along the line of sight 
towards the observer. If $\Nlos > 1$, we would thus expect 
to have several clouds overlapping along the line of sight,
while for $\Nlos < 1$, this number would correspond to the
probability $\Plos$ of one cloud being located along the line 
of sight. The initial depth of the iron K absorption edge of
clouds along the line of sight (assuming that we have at least 
1 cloud located within the line of sight) will add up to
\begin{equation}
\taufecl \sim \cases{ 6.5 \, {\mcl \, \Xfe \over \Abarfourty
\, t_y^2} & if $\Nlos \ge 1$, \cr\cr
11 \, {\mcl \, \Xfe \, \thetaclo \over \Abarfourty \, \ffill 
\, t_y^2} & else \cr}
\label{tauFe}
\end{equation}

The observation of $\taufecl \sim 1.6$ during the first
$\sim 13$~s of GRB~990705 can be used to put parameter
constraints on the progenitor of this GRB. Assuming that
$\Plos \le 1$, this yields
\begin{equation}
\mcl = 16 \, t_y^2 \, {\Plos \, \Abarfourty \over X_{\rm Fe}}.
\label{mcl_tau}
\end{equation}
This relation is plotted for various ratios $\Plos / X_{\rm Fe}$
and $\Abarfourty = 1$ by the solid curves in Fig. \ref{parameterspace}. 

Normalizing the spectrum and luminosity of the ionizing radiation
to the observed spectrum of GRB~990705 during the first $\sim 20$~sec, 
we find a time scale (in the cosmological rest frame of the GRB) 
for complete ionization of initially neutral iron of $\tion (z) 
\sim 2.5 \times 10^{-4} \, t_y^2 \, h_{70}^2 \; {\rm s}$.
For comparison, the recombination time scale can be estimated
to $\trec (z) \sim 9.1 \times 10^{-2} \, T_3^{3/4} \, \Abarfourty 
\, \ffill \, t_y^3 / [(\Zeff / 10)^2 \, \mcl] \; {\rm s}$,
where $T_3$ is the average electron temperature in the clumps in
units of $10^3$~K, and $\Zeff$ is the average effective ion charge
of iron ions in the clumps. 

If recombination is inefficient, then the expression for the 
ionization time scale allows an estimate of the time delay $t_y$ 
from $\tion (z) \sim 6$~s. We find
\begin{equation}
t_y = 63 \, h_{70} \, \left( {\tion [z] \over {\rm s}} \right)^{1/2}. 
\label{ion_timescale}
\end{equation}
This relation is shown for $\tion = 6$~s (which might be the 
appropriate time-scale, in the cosmological rest frame of GRB~990705, 
of the occurrence of the absorption feature) and $h_{70} = 1$ by 
the long-dashed vertical line in Fig. \ref{parameterspace}. 

If we require recombination to compete successfully with 
photoionization, then a comparison of the respective time 
scales yields
\begin{equation}
\mcl \ge 364 \, {T_3^{3/4} \, \Abarfourty \, \ffill \over 
(\Zeff / 10)^2} \, t_y.
\label{ion_equilibrium}
\end{equation}
The most conservative limit can be found if we consider recombination
from fully ionized iron, i.e. $\Zeff = 26$, in which case we find 
$\mcl \ge 54 \, \ffill \, t_y$ for $\Abarfourty = T_3 = 1$. This 
relation is indicated by the shaded regions in the upper-left corner 
of Fig. \ref{parameterspace} for $\ffill = 1$ and $\ffill = 0.1$, 
respectively. If recombination is successfully competing with 
photoionization, then the decline of the Fe~K absorption edge in 
GRB~990705 could be due to Compton heating of the electrons in 
the clouds. The Compton heating time scale in the clouds can be 
estimated through $t_h (z) \sim 3.8 \times 10^{-3} \, t_y^2 \, 
h_{70}^2 \, T_3^{0.1} \; {\rm s}$, which leads to
\begin{equation}
t_y = 16 \left( {t_h [z] \over {\rm s}} \right)^{1/2} \, h_{70}^{-1}. 
\label{t_heating}
\end{equation}
Using $h_{70} = 1$ and $t_h = 6$~s, this relation is plotted by 
the dot-dashed vertical line in Fig. \ref{parameterspace}.

Alternatively, the disappearance of the absorption feature could
be due to sweeping-up of the absorber by the relativistic blast
wave associated with the GRB. Assuming that during the prompt GRB
phase the relativistic blast wave is in the coasting phase, moving
at a constant bulk Lorentz factor $\Gamma = 10^3 \, \Gamma_3$, we
find for the sweep-up time: $t_{sw} (z) \sim x / (2 \Gamma^2 c) 
\sim 8 \times 10^{-2} \, t_y / \Gamma_3^2$~s which, when set equal 
to the $\sim 6$~s time scale of the absorption feature, yields
\begin{equation}
t_y \sim 72 \, \Gamma_3^2
\label{t_sweep}
\end{equation}
For $\Gamma = 100$, 300, and $10^3$, this is plotted by the dotted
vertical lines in Fig. \ref{parameterspace}.

\section{\label{discussion}Discussion}

\begin{figure}
\plotone{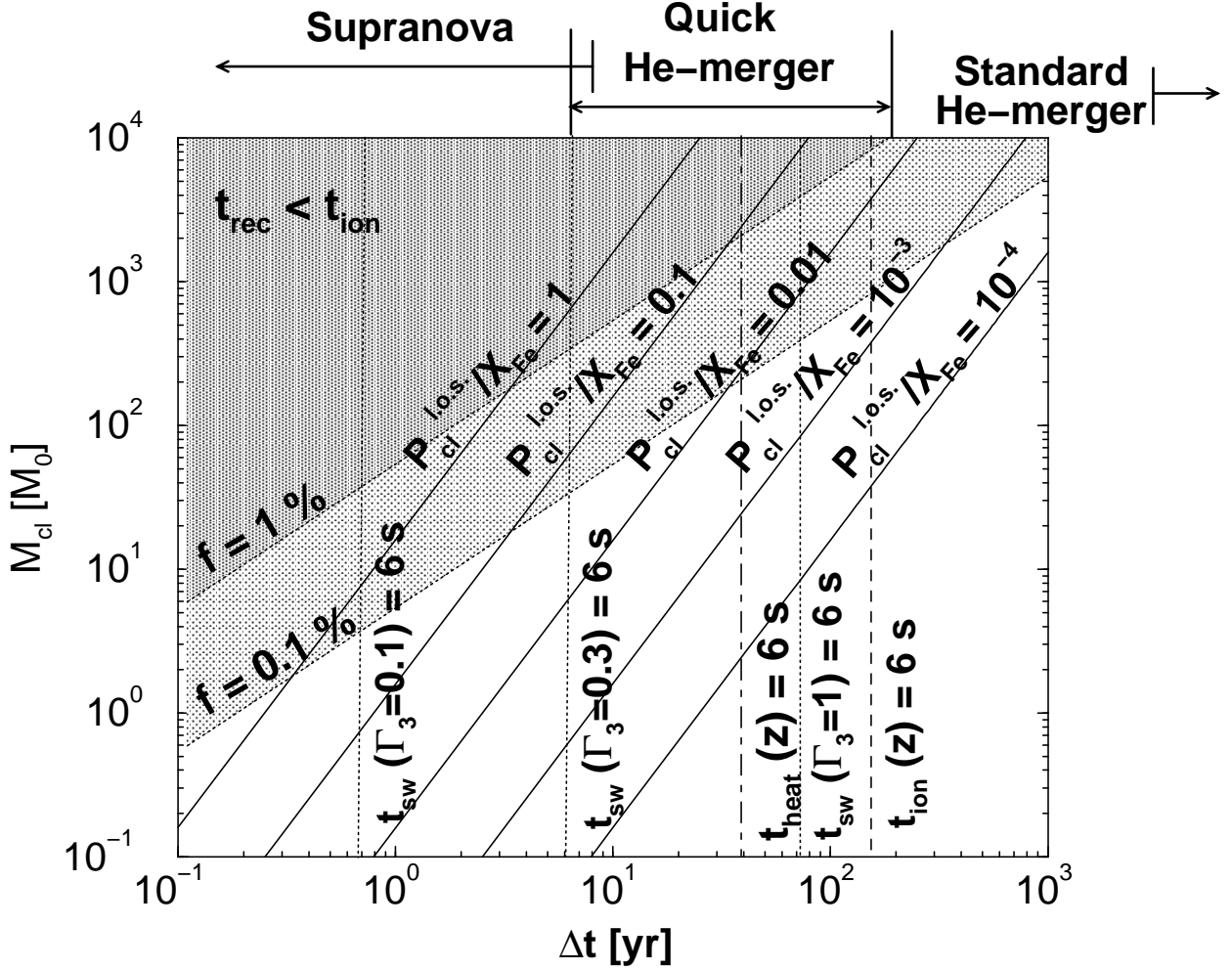}
\caption{Parameter space of supernova ejecta mass $M_{\rm cl}$ 
concentrated in clumps, and time delay $\Delta t$ between the
primary's supernova explosion and the GRB. The solid lines
indicate the condition that an iron K$\alpha$ absorption edge
of the depth observed in GRB~990705 is produced, for various
values of the ratio of the probability $\Plos$ of an absorbing
cloud being located in the line of sight, to the iron enhancement
$X_{\rm Fe}$ w.r.t. standard solar-system values. Constellations
which would give a consistent physical scenario, must either be
located close to the vertical line corresponding to $\tion (z)
= 6$~s (if recombination is inefficient), or within the shaded 
regions in the upper left corner of the plot, which indicates the
condition $t_{\rm rec} \le t_{\rm ion}$ for volume filling factors
of the SN ejecta of 1~\% and 0.1~\%, respectively, 1~year after the
SN.}
\label{parameterspace}
\end{figure}

The transient absorption feature in GRB~990705 places stringent 
constraints on the GRB progenitor parameters. In Fig. 
\ref{parameterspace}, allowed parameter constellations have 
to be located in two regions: a) close to the vertical line
$\tion = 6$~s, or b) in the shaded region in the upper-left
portion of the graph. Case a) corresponds to a pure
photoionization scenario, in which recombination in the
absorber is inefficient; case b) corresponds to a balance
of ionization and recombination. At the same time, ejecta
masses significantly exceeding $\sim 10 \, M_{\odot}$ may be
ruled out. Also, any solution far left of the vertical
line corresponding to $t_{sw} (\Gamma_3 = 0.1) = 6$~s are
implausible since this region of parameter space would
require $\Gamma_0 \ll 100$, in which case $\gamma\gamma$
opacity effects might become significant. Solutions near 
the lower right corner of the diagram would require a very 
strong iron overabundance w.r.t. the standard solar-system 
value, and a small covering fraction of the absorber, 
implying a small chance probability of having an absorbing 
cloud in the line of sight to the observer. 

This leaves only a very limited region of parameter space
to explain the absorption feature in GRB~990705. The
solution at $\Delta t \sim 150$~yr, corresponding to the
no-recombination case, is in principle accessible to 
the quick He-merger scenario, but requires an ejecta
mass of $M_{\rm cl} \gs 10 \, M_{\odot}$, a large
iron overabundance, and a small covering fraction, $\Plos 
/ X_{\rm Fe} \ls 10^{-3}$. This small chance
probability factors in with the already small probability
of $\sim 0.5$ -- 2.5~\% of quick He-mergers among the total
number of expected He-merger GRBs to yield a total probability
of $\ll 1$~\% of observing an absorption feature as seen in
GRB~990705. Consequently, if similar absorption features
will be detected (e.g., by the Swift satellite, currently 
scheduled for launch in 2003) in other GRBs, this may be
a strong indication that at least this type of GRBs might
not be related to collapsars or He-mergers.

The second allowed region of parameter space is located at
time delays of $\Delta t \sim 1$~yr and ejecta masses of
$\sim 10 \, M_{\odot}$. Here, the observed characteristics
of the absorption feature in GRB~990705 can be achieved with 
rather unspectacular values of $\Plos / X_{\rm Fe} \sim 1$.
The only current GRB model which could produce this combination
of pre-GRB supernova ejecta and time delay, seems to be the
supranova model \citep{vs98}. However, it is unclear whether 
the problems mentioned in Section \ref{models} (i.e., the 
rather large baryon contamination and small available energy 
resulting from realistic simulations of such a scenario) can 
be overcome with strong beaming of the GRB ejecta. More 
detailed simulations of the supranova model are needed to 
clarify these issues.

\acknowledgments
The work of MB is supported by NASA through Chandra 
Postdoctoral Fellowship grant PF~9-10007 awarded by the Chandra 
X-ray Center, which is operated by the Smithsonian Astrophysical 
Observatory for NASA under contract NAS~8-39073. C.L.F. is supported
by a Feynman Fellowship at LANL. The work of CD is supported by the 
Office of Naval Research.

\newpage

\newpage

\newpage

\begin{deluxetable}{ccccc}
\tabletypesize{\scriptsize}
\tablecaption{Parameters of the absorber in GRB 990705, from
progenitor-model independent estimates. $M_{\rm Fe}$ is the 
total mass of iron in the absorbing material,$r_c$ is the radius 
of dense clumps containing iron-enriched material, $x$ is their 
average distance from the burst source, and $n_c$ is their average 
density. $\tau_T$ is the Thomson depth of the absorber, and $\Omega$ 
is the solid angle subtended by the absorber as seen from the 
burst source. The iron abundance in the clumps is parametrized 
as $X_{\rm Fe} = 10 \, \Xfe$ times the standard solar-system value, 
and $L_{51}$ is the luminosity of the prompt burst radiation in 
units of $10^{51}$~ergs~s$^{-1}$.
\label{parametertable}}
\tablewidth{0pt}
\tablehead{
\colhead{Quantity} & \colhead{$M_{\rm Fe} \, [M_{\odot}]$} & \colhead{$r_c$ [cm]}   & \colhead{$n_c$ [cm$^{-3}$]} & \colhead{$x$ [cm]}
}
\startdata
Recomb.   \\
ineff.    & $44 \, \Omega$                 & undet.                                      & undet.                                    & $4 \times 10^{18}$                 \\
\noalign{\bigskip}
Photoel.  \\
abs. in   & $0.7 \, \Omega$                & ${7.5 \times 10^{13} \over \Omega \, \Xfe}$ & $10^{11}$                                 & $2 \times 10^{17}$                 \\
dense cl. \\
\noalign{\bigskip}
Resonance \\
scat. in  & $1.3 \times 10^{-3} \, \Omega$ & $2 \times 10^{13} \, \tau_T^2 \, \Xfe$      & ${8 \times 10^{10} \over \tau_T \, \Xfe}$ & $3 \times 10^{16} \, L_{51}^{1/2}$ \\
dense cl. \\
\enddata
\end{deluxetable}


\begin{thebibliography}{}

\bibitem[Amati et al.(2000)]{amati00} 
Amati, L., et al., 2000, Science, 290, 953

\bibitem[B\"ottcher et al.(1999)]{boettcher99}
B\"ottcher, M., Dermer, C. D., Crider, A. W., \& Liang, E. P., 1999, A\&A,
343, 111

\bibitem[B\"ottcher et al.(2001a)]{boettcher01a}
B\"ottcher, M., Dermer, C. D., Amati, L., \& Frontera, F., 2001a, in proc. of 
``Gamma-Ray Bursts in the Afterglow Era II'', in press

\bibitem[B\"ottcher et al.(2001b)]{boettcher01b}
B\"ottcher, M., Dermer, C. D., Amati, L., \& Frontera, F., 2001b, in proc. of 
``Gamma 2001'', AIP proc., in press

\bibitem[Dotani et al.(1987)]{dotani87}
Dotani, T., Hayashia, K., Inoue, H., Itoh, M., \& Koyama, K., 1987, Nature, 
330, 230

\bibitem[Frontera et al.(2000)]{frontera00}
Frontera, F., et al., 2000, ApJS, 127, 59

\bibitem[Fryer \& Woosley(1998)]{fw98}
Fryer, C. L., \& Woosley, S. E., 1998, ApJ, 501, 780

\bibitem[Fryer, Burrows, \& Benz(1998)]{fryer98}
Fryer, C. L., Burrows, A., \& Benz, W., 1998, ApJ, 496, 333

\bibitem[Fryer(2001)]{fryer01}
Fryer, C. L. 2001 Black Holes in Binaries and Galactic Nuclei. Proceedings 
of the ESO Workshop held at Garching, Germany, 6-8 September 1999. Lex 
Kaper, Edward P. J. van den Heuvel, Patrick A. Woudt (eds.), p. 328. Springer

\bibitem[Fryer, Hungerford, \& Warren(2001)]{fhw01}
Fryer, C. L., Hungerford, A., \& Warren, M., 2001, in preparation

\bibitem[Germany et al.(2000)]{germany00}
Germany, L. M., Reiss, D. J., Sadler, E. M., Schmidt, B. P., \& Stubbs, C. W., 
2000, ApJ, 533, 320

\bibitem[Herant \& Woosley(1994)]{hw94}
Herant, M., \& Woosley, S.E. 1994, ApJ, 425, 814

\bibitem[Hughes et al.(2000)]{hughes00}
Hughes, J. P., Rakowski, C. E., Burrows, D. N., \& Slane, P. O., 2000,
ApJ, 528, L109

\bibitem[Hwang, Holt, \& Petre(2000)]{hwang00}
Hwang, U., Holt, S. S., \& Petre, R., 2000, ApJ, 537, L119

\bibitem[Itoh et al.(1987)]{itoh87}
Itoh, M., Kumagai, S., Shigeyama, T., Nomoto, K., \& Nishimura, J., 1987, 
Nature, 330, 233

\bibitem[Kifonidis et al.(2000)]{kifonidis00}
Kifonidis, K., Plewa, T., Janka, H.-Th., \& M\"uller, E., 2000, ApJ, 531, L123

\bibitem[Lazzati et al.(2001)]{lazzati01}
Lazzati, D., Ghisellini, G., Amati, L., Frontera, F., Vietri, M., \&
Stella, L., 2001, ApJ, 556, 471

\bibitem[Matz et al.(1988)]{matz88}
Matz, S. M., Share, G. H., Leising, M. D., Chup, E. L., \& Vestrand, W. T.,
1988, Nature, 331, 416

\bibitem[Ruffert, Janka, \& Schaefer(1996)]{ruffert96}
Ruffert, M., Janka, H.-Th., \& Schaefer, G., 1996, A\&A, 311, 532

\bibitem[Schmidt et al.(1994)]{schmidt94}
Schmidt, B. et al. 1994, AJ, 107, 1444

\bibitem[Sollerman et al.(2000)]{sollerman00}
Sollerman, J., Kozma, C., Fransson, C., Leibundgut, B., Lundqvist, P.,
Ryde, F., \& Woudt P., 2000, ApJ, 537, L127

\bibitem[Spyromilio et al.(1988)]{spyromilio88}
Spyromilio, J., Meikle, W. P. S., Learner, R. C. M., \& Allen, D. A., 
1988, Nature, 334, 327

\bibitem[Sunyaev et al.(1987)]{sunyaev87}
Sunyaev, R. A., et al., 1987, Soviet Astronomy Letters, 13, 431

\bibitem[Turatto et al.(1998)]{turatto98}
Turatto, M. et al. 1998, ApJ, 498, L129

\bibitem[Vietri \& Stella(1998)]{vs98}
Vietri, M., \& Stella, L., 1998, ApJ, 507, L45

\bibitem[Yoshida et al.(1992)]{yoshida92}
Yoshida, A., Murakami, T., Nishimura, J., Kondo, I., \& Fenimore, E. E.,
1992, in: ``Gamma-Ray Bursts --- Observations, Analyses and Theories''
(A93-20206 06-90), p.399

\end{thebibliography}
\end{document}